\documentclass[12pt]{emulateapj}
\newcommand{\beq}{\begin{equation}}
\newcommand{\eeq}{\end{equation}}

\newcommand{\hi}{H{\sc i}~}
\newcommand{\hii}{H{\sc ii}~}
\newcommand{\hia}{H{\sc i}}
\newcommand{\hiia}{H{\sc ii}}
\newcommand{\citei}[1]{\citeauthor{#1} \citeyear{#1}}

\newcommand{\kms}{km ${\rm s^{-1}}$~}
\newcommand{\kmsa}{km ${\rm s^{-1}}$}
\slugcomment{To appear in the Astrophysical Journal}
\begin{document}

\title{Hitting the Bull's-eye: The Radial Profile of Accretion and Star Formation in the Milky Way}

\author{J.~E.~G.~Peek\altaffilmark{1}$^,$\altaffilmark{2}}
\altaffiltext{1}{Department of Astronomy, University of California, Berkeley, CA 94720. goldston@astro.berkeley.edu}
\altaffiltext{2}{Space Sciences Laboratory, University of California, Berkeley, CA 94720}

\begin{abstract}
Ongoing star formation in the Milky Way requires continuous gaseous fuel from accretion. Previous work has suggested that the accretion of dwarf galaxies could provide the needed gas for this process. In this work we investigate whether dwarf galaxy accretion is consistent with the radial profile of star formation observed in the Milky Way, which is strongly concentrated toward the center of the gaseous disk of the Galaxy. Using realistic parameters for the Galactic potential, gaseous halo, Galactic disk, velocities of dwarf galaxies, and effects of drag on stripped gas in the halo, we model the distribution of radii at which dwarf galaxies accrete onto the Galactic disk. We also model the radial distribution of the accretion of gas that cools directly out of the halo by examining the results of recent simulations. We show that dwarf galaxies cannot reproduce the concentration of accretion toward the center of the Galaxy required by star formation. We also show that clouds that cool directly from the halo can reproduce this central concentration, and conclude that this difference is largely due to the discrepancy in absolute specific angular momentum between the two mechanisms. 
\end{abstract}


\section{Introduction}\label{intro}

Observations of young stars show that our Galaxy, like other spiral galaxies, is currently forming new stars out of the interstellar medium. It has been known for some time that this rate of star formation, $ \ge 1 M_\odot$/yr (\citealt{Rana91}; hereafter Rana91), is so fast that it will exhaust all the gas in the star-forming area of the Milky Way (MW), about 2$\times 10^9 M_\odot$ (\citealt{HJK82}; \citealt{Levine2006a}, hereafter LBH06), in a small number of Gyrs. This timescale is much shorter than the age of the MW ($> 10$ Gyrs), which leads us to conclude that we are either in a special moment in the evolution of the MW or that the MW is continuously gaining fuel for star formation.

Work by \citet{CMG97} (hereafter CMG97), among others (see \citealt{Larson72} for an early discussion), have bolstered the view that the MW is currently accreting new gas to form stars. These works showed that the distribution of metallicities of stars in the MW can be reproduced by invoking two epochs of gas infall: one intense, early epoch and one slower, ongoing epoch. These works also showed that star formation in the absence of the continuous accretion of fresh, relatively unenriched gas cannot generate the correct distribution of metals in the stars we observe: if all the gas had been available for star formation since z=1 we encounter the ``G-dwarf problem,'' wherein models predict too many long-lived stars with very low metallicities compared to observations.

In an extension of this type of stellar population metallicity modeling, \citet{CMM08} (hereafter CMM08) showed that if gas accretion traces the accretion of smaller dark matter (DM) satellites (i.e., dwarf galaxies) in cosmologically simulated galaxies, the MW distribution of metals in stars can also be reproduced. CMM08 also showed that this claim is only true in simulated Galaxies with relatively quiescent recent merger histories, i.e.,\ with all merger ratios below 1:20 in the last 8 Gyrs (since z = 1). These galaxies only accreted relatively small dwarfs during that time, consistent with the slower, ongoing accretion epoch from CMG97. \citealt{Stewart07} have shown that about 40\% of MW-sized halos in simulations of standard $\Lambda$CDM cosmology have merger histories that meet or exceed this recent merger quiescence criterion. This is to say that the constraints from CMM08 put the MW in the 40\% most quiescent MW-sized haloes, and are therefore not necessarily unreasonable.

\begin{figure}
\begin{center}
\includegraphics[scale=0.8, angle=0]{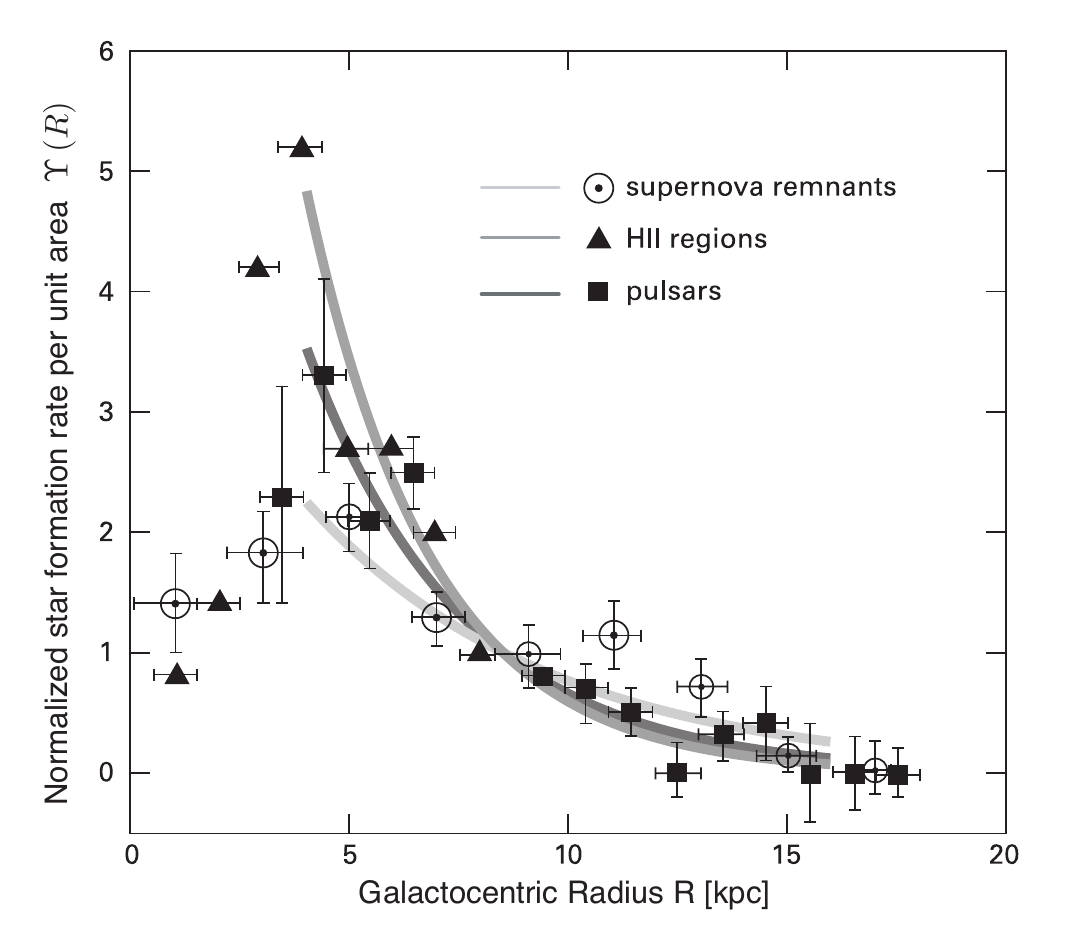}
\caption{The observed MW star formation rate per unit area normalized to the solar circle as measured by supernova remnants, \hii regions, and pulsars. Also shown are our fits to Equation \ref{expo} over the 4 kpc $< R <$ 15 kpc range for each tracer. This plot was adapted with permission from \citet{SP05}.}
\label{stahler}
\end{center}
\end{figure}

To do this modeling CMM08 had to assume that the accreted gas would be distributed within the Galaxy to match the observed radial distribution of star formation, as the DM-only cosmological simulations did not trace the evolution of the gas itself. By tracing the products of star formation (\hii regions \citep{GM82}, Supernova remnants \citep{LW89}, and pulsars \citep{Lyne85}), it has been shown that star formation in the MW is a strong function of Galactic radius, peaking at a Galactic radius $R \simeq 4$ kpc, and tapering off to $R \simeq 15$ kpc. The rate of this decrease varies depending upon the star formation tracer employed, but it can be fit by an exponential function, 
\beq\label{expo}
\Upsilon \left( R \right) = e^{\beta \left(R-R_\odot\right)},
\eeq
where $\Upsilon$ is the star formation rate per unit area normalized to the solar circle (R = 8.5 kpc) and $-0.35 {\rm~kpc}^{-1} \le \beta \le -0.18 {\rm~kpc}^{-1}$ (see Figure \ref{stahler}). 

It was suggested by \citet{MB2004} that a massive gaseous halo ($M_{\rm halo} \simeq 10^{11} M_\odot$) around the MW could also be a source of fuel for ongoing accretion onto the Galactic disk. In this theory, the hot ($T \sim 10^6$ K) halo unstably cools (see \citealt{Field65}, \citealt{ML90}) to form overdense clouds that then fall onto the Galactic disk. Simulations have since confirmed that a massive halo can generate such clouds (e.g., \citealt{SL06}; \citealt{Kaufmann05}; and \citealt{Kaufmann08}, hereafter KBMF08) and that this cooling halo material is consistent with the observed HVCs in the Galactic Halo, excepting the Magellanic Stream (\citealt{PPS-L08}; hereafter PPS-L08). We therefore investigate these cooling halo clouds, alongside dwarf galaxies, as an alternative source of fuel for star formation in the MW. Indeed, it is not known which of these gas accretion mechanisms, dwarf galaxy stripping or the cooling of a massive halo, dominates accretion onto spirals in the local universe (e.g., \citealt{FB08}).

In this work we examine the accretion of gas from small dwarf galaxies and cooling halo clouds in terms of their consistency with the observed radial distribution of star formation in the MW. In particular, we make a two conjectures about the radial movement of gas in the Galaxy (\S \ref{radmot}), which we use to compare the observed radial distribution of star formation and the modeled radial distribution of accretion. We construct a model of Galactic accretion, which is dependent upon a variety of physical parameters of the Galaxy (\S \ref{gal}) and the accreting gas (\S \ref{dgal}). In \S \ref{gal} we discuss the relevant parameters of the Galaxy, including the extent of gas in the Galactic disk (\S \ref{galdisk}), the gravitational potential of the MW (\S \ref{gpot}) and the structure of the Galactic gaseous halo (\S \ref{gashalo}). In \S \ref{dgal} we discuss the kinematic and physical characteristics of dwarf galaxies impinging on the MW (\S \ref{kindwarf} and \S \ref{intchar}), methods by which gas can be stripped from these galaxies (\S \ref{strip}), and the characteristics of such gas once it is stripped (\S \ref{sgstruct}). In \S \ref{meth} we describe our method for modeling gas accretion onto the Galactic disk. We discuss our findings in \S \ref{results} and conclude in \S \ref{concl}.

\section{Radial Motion of Gas Within the Galactic Disk}\label{radmot}

As the MW presents a large area to accreting material (see \S \ref{galdisk}), yet only forms stars towards the center, it is natural to ask whether gas accreted at one radius in the disk can propagate inward to form stars at another radius. Unfortunately, we are not provided with significant observational constraints. Given the radial surface density profile of gas in the MW (LBH06), to fuel central star formation with gas from the Galaxy beyond the star-forming radius, an inflow velocity of order only 5 \kms is needed. This fueling is too slow to detect in observations of \hi in the MW. \citet{WBB04} investigated inflow in other spiral galaxies, but were unable to put constraints on the inflow rate strong enough to rule out this relatively slow feeding.

From a theoretical standpoint, to move material inward angular momentum must be redistributed. Indeed, it is energetically favorable for most of the disk gas to give its angular momentum to a small amount of material through some global mechanism. In this way most of the gas propagates inward while a small amount moves ever further from the center. The standard mechanism invoked for this angular momentum redistribution is spiral density waves (e.g., \citealt{L-BK72}, hereafter L-BK72). To determine whether the torques generated by spiral density waves can generate significant inflow in the disk we define the inflow timescale, $t_{\rm inflow} \equiv L/\tau$, where $L$ is the angular momentum in the disk and $\tau$ is the torque from spiral arms. We let $L = \Sigma R^3 v_\phi$, and we use the derivation of spiral torque from \citep{BT08} for small pitch-angle spiral arms,
\beq
\tau = \frac{\pi^2mRG\left(\delta\Sigma\right)^2}{k^2}
\eeq
where $\delta\Sigma$ is the change in surface density in the spiral arms, $k$ is the radial wavenumber (in rad$^{-1}$) and $m$ is the number of spiral arms. Inserting values for all of these parameters reasonable for the outer disk of the MW (see LBH06 for gaseous surface density maps) we find
\beq
t = 1.2 \times 10^{10} {\rm yr} \left(\frac{4}{m}\right)\left(\frac{Rk}{2\pi}\right)^2\left(\frac{\Sigma}{4 M_\odot {\rm pc}^{-2}}\right)\left(\frac{4 M_\odot {\rm pc}^{-2}}{\delta\Sigma}\right)^2,
\eeq
approximately the age of the Universe. This is much longer than the timescale required for fueling in the disk, $\sim 10^9$ years. 

Of course, we cannot conclusively claim that gas does not move radially from the outer disk to the areas of star formation. While torques from gaseous spiral arms do seem unlikely to be able to move gas on the relevant timescale, without empirical evidence we do not know whether some other mechanism might be at work. Indeed, other mechanisms have been proposed to induce significant radial inflow. We therefore consider two criteria, each of which could be true or false, regarding the movement of gas in the disk: the ``weak no-flow criterion'' and the ``strong no-flow criterion.'' The weak no-flow criterion is that gas accreted beyond the maximum star-forming radius ($R > 15$ kpc) of the MW does not move into the star forming region. The strong no-flow criterion is that gas accreted at any radius in the MW outside 4 kpc forms stars only at the radius at which it lands. Since the MW bar, which could extend out to 4 kpc, can be a very efficient mechanism for the redistribution of angular momentum (e.g. \citealt{BS91}, L-BK72), we make no assumptions about the region interior to 4 kpc, and omit it from our analysis. 

We use these two criteria to test our models of Galactic accretion. If a model of MW accretion puts too much material in the outer disk, it will not meet the weak no-flow criterion. If a model of the MW accretion distributes gas along the disk with a surface density inconsistent with the observed profile of star formation, it will not meet the strong no-flow criterion. We note that the strong no-flow criterion is very stringent, and certainly not required by spiral torque theory. We include this as a criterion only to show whether models do or do not require movement within the star-forming disk. We describe the quantitative constraints on accretion that stem from these two criteria below.

We know the amount of gas in the MW beyond the maximum star forming radius dominates the mass of stars in that region, $M_{HI, {\rm out}} \sim 1.7 \times 10^{9} M_\odot$ (LBH06), and that the current star-formation rate in the MW is $ \ge 1 M_\odot$/yr. Since all accretion histories that are consistent with thin-disk stellar metallicities have a broadly declining accretion rate over the last 8 Gyrs \citep{CMG97}, we can take a lower limit of the total star formation since $z \sim 1$ within the MW to be $8 \times 10^{9} M_\odot$. A given generation of stars will return  48\% of its gas to the ISM \citep{Rana91}, which, after many generations, means that a unit of accretion will yield about 2 units of star formation. We therefore only require half the star formation rate to be represented by accretion onto the MW during this time, $M_{\rm acc} = 4 \times 10^{9} M_\odot$. If the amount of gas a model places beyond $R = 15$ kpc is too high compared with the amount it places within $R = 15$ kpc to match these observational constraints, the model is inconsistent with the weak no-flow criterion: gas must propagate inwards in the disk. We define this limiting fraction as $a_{\rm out} \equiv M_{HI, {\rm out}}/M_{\rm acc} = 0.44$.

Given the observations of star-formation tracers (Figure \ref{stahler}), we were able to parameterize the radial profile of star formation within the disk with Equation \ref{expo}. For a model of accretion onto the MW to be consistent with the strong no-flow criterion, the radial profile of accretion in the model must fall within the range of these radial profiles of star formation. If at some place in the disk the accretion were to lag behind the star formation, the star formation would have ceased; if the star formation were to lag behind the accretion we would see large gas reservoirs at that radius. Since neither of these is seen, any model which has a radial accretion profile with a $\beta < -0.35 {\rm~kpc}^{-1}$ or $\beta > -0.18 {\rm~kpc}^{-1}$ is inconsistent with the strong no-flow criterion.

\section{The Galaxy}\label{gal}

	\subsection{The Scale of the Galactic Disk}\label{galdisk}

The extent of the Galaxy disk can be measured in terms of DM, stars, or gas. In this case we wish to determine the extent of the Galactic disk in terms of the ability to accrete (arrest) gas. Since infalling gas could pass by a disk of stars or of DM, but not of gas, it is the extent of the gaseous disk that is relevant to our disk size measurement. LBH06 show through observations of \hi that the MW disk has a surface density above 0.1 $M_\odot$ pc$^{-2}$, or 3 $\times 10^{19} {\rm cm}^{-2}$ in hydrogen atom column density within $R_D \simeq$ 30 kpc. This is the relevant column density to arrest the infall of accreting material, as most accreting gas is expected to be near this column (see \S \ref{sgstruct}). If we choose instead a minimum surface density to arrest accreting gas of $10^{20} {\rm cm}^{-2}$, we find a maximum radius of $\sim$ 25 kpc, which leaves the qualitative conclusions of this work unchanged. Therefore, in our modeling, we take the Galactic disk to be capable of accreting gaseous material out to $R_D = 30$ kpc

	\subsection{The Gravitational Potential of the Galaxy}\label{gpot}

To understand the process of accretion in the MW one must first parameterize the dominant force acting on accreting material: gravity. The gravitational potential of the Galactic DM halo can be modeled by examining the movement of stars and gas in various positions in the MW and MW halo. We use the model from \citet{Wolfire1995}, consisting of a Miyamoto-Nagai disk potential, a spherical bulge potential and a logarithmic DM halo potential. These three potentials are added together and constitute the total Galactic potential we use in our models of accretion and the structure of the gaseous halo. 

	\subsection{The Gaseous Halo}\label{gashalo}

In this work, when we refer to the gaseous Galactic halo, we are referring to the gas that occupies the volume surrounding the MW out to approximately the virial radius ($R_{vir} \sim 250$ kpc), where it is assumed to merge with the inter-galactic medium. It is this halo gas that will strip gas out of dwarf galaxies and generate drag on these stripped clouds. It is therefore crucial to determine the extent and character of the gaseous halo.

Observations the soft X-ray background (e.g., \citealt{KS2000}), the pressure confinement of high-velocity clouds (HVCs; e.g., \citealt{Wolfire1995}), and absorption of high ions in X-ray spectra (e.g., \citealt{FMCW06}) suggest the the existence of such a halo. Constraints on the X-ray emission measure from the halo (4.7 $\times 10^{-3}$ ${\rm cm}^{-6}$ pc) and temperature constraints from both the soft X-rays and absorption by high ions ($10^6$ K $\le T_{\rm halo} \le 3 \times 10^6$) allow us to construct a simple hydrostatic, isothermal halo within the Galactic potential that depends upon only the temperature of the halo. Unfortunately, this relatively narrow range in temperature can significantly affect the mass of the halo and density of the halo beyond $\sim 30$ kpc. We therefore use one of two distinct models of the halo in each of our runs: a ``low-mass'' halo, with T = $10^6$ K and $M_{halo} \simeq 10^9 M_\odot$, and a ``massive'' halo with T=$3 \times 10^6$ K and $M_{halo} \simeq 10^{11} M_\odot$. We note that while this is certainly a simplistic formulation of the halo, the massive halo is consistent with the density structure of the simulated halo described in \citet{KBMF08} to within a factor of 2.

\section{Dwarf Galaxies}\label{dgal}

	\subsection{Kinematic Characteristics of Infalling Dwarf Galaxies} \label{kindwarf}

To determine the distribution of gas deposition on the Galactic disk by dwarf galaxies we must first determine their initial conditions as they fall into the Galactic potential. We would prefer to determine this by observational evidence, but the observed dwarfs around the MW may not be a good representation of typical dwarfs entering the influence the MW for the first time. Dwarfs on a ``collision course'' with the core of the MW disk may only be observed once near the virial radius, whereas dwarfs that avoid such collisions may be observed on many subsequent orbits. More generally, the ellipticity and velocity of dwarf galaxy orbits will determine the amount of time a given galaxy spends near the virial radius. Additionally, we do not have a full census of dwarf galaxies at the MW virial radius (see \citealt{Koposov07}). 

Instead of using observational data, we use radial and azimuthal velocity distributions determined from DM simulations. \citet{Benson05} compiled data from a number of different DM cosmological simulations to determine the distribution of radial and tangential velocities of satellites joining larger halos at the virial radius, $f\left(v_r, v_\phi \right)$. He found that $f\left(v_r, v_\phi \right)$ is relatively constant over a wide range of primary halo masses and cosmic time. We draw our sample of dwarf galaxy velocities from this probability distribution function to test our accretion hypotheses. We note that this distribution of subhalo velocities applies both to the dwarf galaxies we wish to examine and and DM haloes that may not contain galaxies.

	\subsection{Mass of Infalling Dwarf Galaxies}\label{intchar}

Dwarf galaxies around the MW and the local group have a large range of sizes, starting with the Large Magellanic Cloud (LMC) with $M \simeq 2 \times 10^{10} M_\odot$ and going beyond our observational limits at near $10^{6} M_\odot$ (\citealt{Koposov07}, \citealt{SG07}). Since our observations are rather incomplete, we cannot determine the relative importance of large and small satellites in fueling accretion onto the MW. While both extended Press-Schecter theory and simulations show that, per decade, massive satellites contribute most to the accretion rate of MW-sized Galaxies (e.g., \citealt{FM08}, \citealt{Stewart07} and references therein), they also show that much of the accretion is generated by the many decades of smaller mass satellites. \citet{Madau08} show that sub-halos in very high resolution DM simulations have equal DM mass per decade, but they do not address the question of halo accretion directly. Given this uncertain situation we cannot limit our investigation to only massive or low-mass satellites, as either may be the dominant source of Galactic fuel. To test this broad range of dwarf galaxy masses, we run each model with a single dwarf galaxy mass parameter, which we allow to vary over a large range from model to model. As we will see in the next section, dwarf galaxy mass determines when the gas becomes stripped from the stars.

	\subsection{Gas Stripping and Drag in Dwarf Galaxies}\label{strip}

Until gas is stripped out of dwarf galaxies it has little chance of entering the disk of the Galaxy, as dwarf galaxies will retain the majority of their angular momentum, and typically have perigalactions $> 30$ kpc. Gas can be removed from galaxies (without being destroyed) by either ram-pressure stripping or tidal stripping. In ram-pressure stripping, the pressure exerted by the halo gas ``head-wind'' on the dwarf galaxy gas exceeds the gravitation force per unit area on the gas from the dwarf galaxy itself. In tidal stripping, the differential force on two ends of the dwarf galaxy from the MW gravity tears the galaxy apart. This is facilitated by generally larger size of the gaseous component of gas-rich dwarf galaxies as compared to the stellar component. In the case of the mergers of interest, we will neglect tidal stripping as a method for removing gas from dwarf galaxies and focus only on the effects of ram-pressure stripping. We will verify the inconsequentiality of this assumption on our results in \S \ref{results}.

It is important to realize that some or all of this stripped gas could eventually shock, become hot and merge with the gaseous halo. We cannot distinguish in our modeling between (a) gas that was originally part of a dwarf galaxy, merged with the halo, and later `re-cooled' via the process described in \S \ref{intro} and (b) gas that cools from the halo having never been a part of a dwarf galaxy. Given this, we allow the dwarf galaxy accretion scenario the `benefit of the doubt' and assume that material does not dissipate into the Galactic halo. This allows us to ask whether dwarf galaxies can fuel Galactic accretion in the absence of a massive, cooling halo.

To determine when a dwarf galaxy becomes stripped we use the criterion established in \citet{GG72}. Their stripping criterion can be formulated as
\beq\label{GG}
\zeta \equiv \frac{1}{3}n_{\rm dwarf}\sigma^2 < n_{\rm halo}v^2,
\eeq
where $n_{\rm halo}$ is the number density of the halo at the location of the dwarf, $v$ is the velocity of the dwarf with respect to the halo gas, $\sigma$ is the velocity dispersion of the stars in the dwarf galaxy, and $n_{\rm dwarf}$ is the typical number density of gas in the dwarf galaxy (see \citealt{GP08}). As we stated in \S \ref{intchar}, there is a large range of possibly relevant dwarf galaxy mass, so we choose three scenarios to test: one in which most satellites are so small that they immediately exceed the criterion in Equation \ref{GG} upon entering the halo ($\zeta = 0$ (\kmsa)$^2$ cm$^{-3}$), one in which most dwarf galaxies have $\sigma = 10$ \kms and $n_{\rm dwarf} = 1 {\rm~cm}^{-3}$ ($\zeta = 33$ (\kmsa)$^2$ cm$^{-3}$), and one in which the dwarf galaxies never become stripped ($\zeta = \infty$ (\kmsa)$^2$ cm$^{-3}$). These extremal cases are not intended to represent reality, but rather act as tests of the sensitivity of the model results to the nature of gas stripping.

Once gas is stripped from the dwarf, ram pressure drag from the gaseous halo of the MW can become a significant force on the resulting cloud. To model the effect of this drag on the cloud we use the simple ram-pressure drag expression, 
\beq
F_d = \frac{1}{2}\rho_{\rm halo} v^2 C_d A,
\eeq
where $A$ is the area of the cloud and $C_d$ is the drag coefficient. Following \citet{BD1997}, we assume $C_d =$ 1. This equation is equivalent to
\beq\label{drag}
a_d = \frac{n_{\rm halo} v^2 C_d}{2 N_H},
\eeq
where $a_d$ is the acceleration of the cloud and $N_H$ is the hydrogen column number density of the cloud.

	\subsection{The Structure of Stripped Gas}\label{sgstruct}

In the Galactic halo we have evidence of stripped gas from dwarf galaxies in observations of the Magellanic Stream. The Magellanic Stream is a long chain of gas, detected primarily in \hi as HVCs. HVCs are seen all over the sky, but the Magellanic Stream is the only undoubtable case of dwarf galaxy stripping around our Galaxy. Therefore we we use the Magellanic Stream to determine the typical column density of material that has been stripped from dwarf galaxies. 

\begin{figure}
\begin{center}
\includegraphics[scale=.30, angle=0]{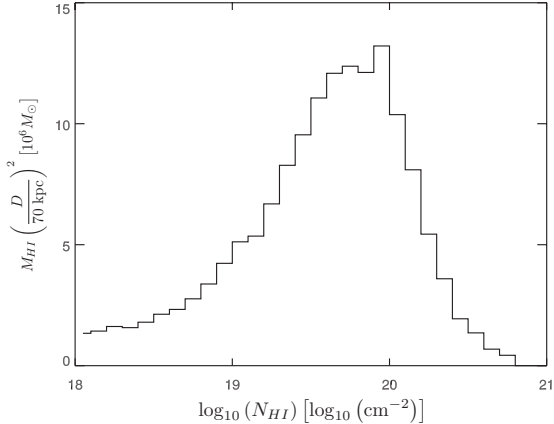}
\caption{The distribution of mass of the Magellanic Stream in \hi by logarithmic column density, assuming that the Stream lies 70 kpc from the observer. Data are from \citet{Putman03a}. The mass-weighted column density of the Stream peaks near $10^{20}$ cm$^{-2}$.}
\label{magstream}
\end{center}
\end{figure}

Unfortunately, there seems to be no specific preferred column density in surveys of the Magellanic Stream or HVCs in general \citep{Wakker2004}. HVC columns begin above $10^{20}$ ${\rm cm}^{-2}$, but have a lower bound that is determined by the sensitivity of the observations. For the question of accretion, though, we are not concerned with the typical column density, but rather the typical \emph{mass-weighted} column density:
\begin{eqnarray}
\langle N_H \rangle &=& \frac{\sum_{i=1}^{M}N_{H,i}m_i}{\sum_{i=1}^{M}m_{i}} = \frac{\sum_{i=1}^{M}N_{H,i}N_{H,i}\Omega D^2\mu m_p}{\sum_{i=1}^{M}N_{H,i}\Omega D^2\mu m_p} \\ 
&=&  \frac{\sum_{i=1}^{M}N_{H,i}^2}{\sum_{i=1}^{M}N_{H,i}}, \nonumber
\end{eqnarray}
where $i$ ranges over resolution elements, $M$ is the number of such elements, $N_{H, i}$ is the column density of each element, $m_i$ is the  mass of the cloud at each element, $\Omega$ is the solid angle subtended by such an element, and $D$ is the distance to the cloud. This mass-weighting effectively diminishes the significance of the long tail towards low column density, which contributes little to the overall accretion rate. For the observations of the entire Magellanic Stream by \citet{Putman03a} at 15$^\prime$.5 resolution, $\langle N_H \rangle = 10^{19.6} {\rm cm}^{-2}$, with most of the mass of the Stream residing between $10^{19.4} {\rm cm}^{-2}$ and $10^{20} {\rm cm}^{-2}$ (see Figure \ref{magstream}). 

We note that due to resolution effects the measured $\langle N_H \rangle$ is a strict lower limit. This is to say that it is possible that the Magellanic Stream is made up of many, higher column density clouds unresolved by observations, which would yield a higher true $\langle N_H \rangle$, although HVC absorption-line observations of distant quasars show that the column densities are relatively converged at a resolution of 10$^\prime$ \citep{Wakker01a}.

We may also want to consider gas accretion from DM subhaloes that have not formed stars (dark galaxies). Modern simulations predict that $\sim$70\% of the DM accreted in subhaloes will have masses above 10$^6 M_\odot$ and that there are of order 15000 such haloes surrounding the MW, far more than are observed (\citei{Springel08}; e.~g.~\citei{SG07}). Were these haloes to contain condensed gas, but not stars, they could be a dominant accretion mechanism of the Galaxy, and, due to their size and starlessness, could have significantly different gas distributions than are observed in the MS. \citet{Westmeier2005} conducted observations of outskirts of Andromeda and ruled out gas in such dark galaxies above $6\times10^4 M_\odot$ in a region that should contain dozens of galaxies with $M_{DM} > 10^6 M_\odot$ and many galaxies with $M_{DM} > 10^7 M_\odot$. This puts the gas fractions of possible dark galaxies below $10^{-3}$, far below the typical $\sim 10^{-1}$ of standard dwarf galaxies \citep{Mateo98}, implying that if such subhaloes do exist they do not contribute significantly to accretion of condensed gas onto the MW. By contrast, the lowest mass known gas-bearing galaxy, Leo T, contains significant dense \hia, with most mass residing at column densities above $10^{20} $cm$^{-2}$. We therefore do not further consider the effects of the accretion of gas via dark galaxies.

This discussion has only addressed the question of stripped neutral gas, mostly because observations of diffuse \hi are much easier to conduct that of diffuse \hiia. Observations in H$\alpha$ of the Magellanic Stream by \citet{Putman03} have shown a significant amount of warm \hii gas exists in this system, although estimating the mass or column density of warm \hii from these measurements is difficult. Simulations conducted by \citet{B-HSAM07} reproduce the observed H$\alpha$ in the Stream. Their results show that the typical mass-weighted column density of warm \hii is above $10^{20}$ cm$^{-2}$ and exceeds that of the \hia. We therefore tentatively conclude that in the case of gas stripped from dwarf galaxies, the contribution from warm \hii can only increase the typical column density from the value measured from \hia, and we therefore consider our $N_H$ estimate from the \hi in the Magellanic Stream to be a lower limit. 

\begin{deluxetable}{cc}
\tablecaption{Parameter list}
\tablewidth{0pt}
\tablehead{
\colhead{parameter} & \colhead{value / reference}  \\
%\hline
}
\startdata
$R_D$ & 30 kpc \\
$R_{vir}$ & 250 kpc \\
$\Phi_{MW}$ & \citet{Wolfire1995} \\
$f\left(v_r, v_\phi \right)$ & \citet{Benson05} \\
$N_H$ & $> 10^{19.6}$ cm$^{-2}$ \\
$M_{halo}$ & $10^9 M_\odot$, $10^{11} M_\odot$ \\
$T_{halo}$ & $10^6$ K, $3 \times 10^6$ K \\
$C_d$ & 1
\enddata
\tablecomments{ A summarized list of parameters used in our models or the reference to the functions used.}
\label{parms}
\end{deluxetable}

\section{Methods}\label{meth}

Given these parameters of the disk, halo, and satellites of the MW (summarized in Table \ref{parms}), we construct a first-order finite-difference code to model the orbits of dwarf galaxies and clouds within the halo. The code evolves the position and velocity of a large number of objects, each representing a cloud or a dwarf galaxy around the MW. Galaxy objects are subject to only the force of gravity, $\nabla \Phi$, from the MW. Cloud objects are subject to both gravity and ram pressure drag (as in Equation \ref{drag}). Note that this code does not capture any of the more complex physical processes that are modeled in modern hyrdrodynamical simulations. We believe that our simplified treatment is an appropriate model for this unstudied question.

In our model runs of dwarf galaxy accretion, $10^4$ galaxy objects are started at random positions at the virial radius with velocities drawn from the distribution from \citet{Benson05}. These galaxy objects orbit until the stripping criterion is met (Equation \ref{GG}), at which point we convert the galaxy object into a cloud object. We follow the orbits of these cloud and galaxy objects until they cross the disk of the galaxy within 30 kpc, or until 8 Gyrs pass in the the model. 

In our model runs of condensing cloud accretion we use the velocities and positions from clouds found in simulations from PPS-L08 and KBMF08 as our initial conditions, and evolve them as cloud objects as we would in the dwarf galaxy accretion model. Note that we do not use any information from these simulations other than the positions and velocities of the clouds selected at an arbitrary time-slice ($0 \le z \le 1$) to give a comparable treatment to the dwarf galaxy model. 

The results of the model runs are radii at which each object lands in the the Galactic disk. Since the total number objects and masses of the objects are irrelevant to the physics of the model, we do not generate any constraints related to the overall accretion rate of the disk. The overall accretion rate is a completely free parameter in our model; readers interested in the consistency between the MW star-formation rate and the overall accretion rate of cooling halo clouds and dwarf galaxies should refer to PPS-L08 and GP09, respectively. Instead, the primary results of these model runs are the relative accretion rates as a function of radius. We use this accretion as a function of radius to test whether each model meets the weak no-flow criterion ($a_{\rm out} < 0.44$) and whether it meets the strong no-flow criterion ($-0.35 {\rm~kpc}^{-1} \le \beta \le -0.18 {\rm~kpc}^{-1}$). 

Given the uncertainties in the halo profile (\S \ref{gashalo}), the relevant cloud column density $N_H$ (\S \ref{sgstruct}), and the stripping criterion $\zeta$ (\S \ref{strip}), we have a variety of runs for each combination of these criteria. To simplify the interpretation of the results, for each run we choose single values for $N_H$ ($10^{19.6}$ or $10^{20.6}$ cm$^{-2}$) and $\zeta$ (0, 33, or $\infty$ (\kmsa)$^2$ cm$^{-3}$), which we apply to all clouds and galaxies in a given run. Results are shown in Table \ref{runs} and discussed below.

\begin{deluxetable*}{ccccc|ccc}
\tablecaption{Model run results}
\tablewidth{0pt}
\tablehead{
\colhead{Run\#} & \colhead{method} & \colhead{halo mass} & \colhead{$\zeta$} & \colhead{log $N_{H, {\rm cloud}}$} & \colhead{$a_{\rm out}$} & \colhead{$\beta$} & \colhead{$f_{\rm accr}$}\\
\colhead{} & \colhead{}& \colhead{}&  \colhead{(\kms)$^2$ ${\rm cm}^{-3}$}& \colhead{log ${\rm cm}^{-2}$}  &\colhead{} & \colhead{1/ kpc} &\colhead{}
}
%\hline
%\hline
\startdata
1 & dwarfs & low-mass & $\infty$ & 19.6 & 2.9 &$-0.03 \pm 0.02 $ & 0.09 \\
2 & dwarfs & low-mass & $\infty$ & 20.6 & 3.1 &$ 0.02 \pm 0.02 $  & 0.09 \\
3 & dwarfs & low-mass & 33 & 19.6 & 2.7 &$-0.01 \pm 0.03 $ & 0.09\\
4 & dwarfs & low-mass & 33 & 20.6 & 2.9 &$ 0.05 \pm 0.03 $  & 0.08\\
5 & dwarfs & low-mass & 0 & 19.6 & 3.9 &$ 0.10 \pm 0.01 $  & 1.00\\
6 & dwarfs & low-mass & 0 & 20.6 & 4.0 &$-0.00 \pm 0.02 $  & 0.19\\
7 & dwarfs & massive & $\infty$  &19.6 & 2.5 & $ 0.02 \pm 0.02 $  & 0.09 \\
8 & dwarfs & massive & $\infty$ & 20.6 & 2.8 & $-0.02 \pm 0.02 $  & 0.09\\
9 & dwarfs & massive & 33 & 19.6 & 3.0 &$-0.07 \pm 0.02 $ & 0.09\\
10 & dwarfs & massive & 33 & 20.6 & 2.8 & $-0.01 \pm 0.02 $  & 0.09\\
11 & dwarfs & massive & 0 & 19.6 & 4.5 & $ 0.10 \pm 0.01 $  & 1.00\\
12 & dwarfs & massive & 0 & 20.6 & 4.5 & $-0.02 \pm 0.02 $  & 0.19\\
13 & PPS-L08 & massive & N/A & 19.6 & 0.07 & $-0.20 \pm 0.02 $  & 1.00\\
14 & PPS-L08 & massive & N/A & 20.6 & 0.19 & $-0.13 \pm 0.03 $  & 1.00\\
15 & KBMF08 & massive & N/A & 19.6 & 0.13 & $-0.27 \pm 0.02 $  & 1.00 \\
16 & KBMF08 & massive & N/A & 20.6 & 0.26 & $-0.27 \pm 0.02 $  & 1.00\\
\enddata
%\hline
\tablecomments{The input parameters for each of the models we run to the left of the divider, with the results at the right of the divider. The large number of models is required to vary each model parameter independently. The method column denotes whether we generate the clouds from dwarf galaxies (as in \S \ref{kindwarf}) or from one of the two cooling halo simulations. The $\zeta$ column denotes at what point gas is stripped from dwarf galaxies, as described in \S \ref{strip} and Equation \ref{GG}. The log $N_{H, {\rm cloud}}$ column denotes the assumed column density of the clouds. We measure the steepness of the normalized accretion rate per unit area with radius ($\beta$, see Equation \ref{expo}) and the ratio of material accreted beyond 15 kpc to that accreted within, $a_{\rm out}$. We also measure the fraction of clouds or galaxies accreted onto the disk within 8 Gyrs, $f_{\rm accr}$.
}
\label{runs}
\end{deluxetable*}

\section{Results}\label{results}

In Table \ref{runs} we show the results from the dwarf galaxy and cooling cloud accretion models. For each model we show two parameters. The first parameter, $a_{\rm out}$, is the ratio of material accreted beyond 15 kpc to that of the material accreted within 15 kpc, which determines whether a model meets the weak no-flow criterion. The second parameter is the coefficient of the best exponential fit to the accretion rate radial profile between 4 and 12 kpc ($\beta$, as in Equation \ref{expo}), which determines whether a model meets the strong no-flow criterion. \emph{All 12 dwarf galaxy accretion models fail to meet either the weak or the strong no-flow criteria.} It is also interesting to note that only in the simulations with $\zeta = 0$ (\kmsa)$^2$ cm$^{-3}$ and $N_H = 10^{19.6} {\rm cm}^{-2}$ did all of the gas accrete onto the disk within 8 Gyrs. This result is consistent with the supposition that stripping and drag are crucial to gas accretion from satellites. We also note here that adding further stripping through tidal effects cannot change this result, as even using immediate stripping the modeled accretion does not meet the criteria.

By stark contrast, all of the models generated from simulated data of cooling haloes meet \emph{both} the weak and the strong no-flow criteria. In a cooling-halo scenario, therefore, no radial gas movement within the disk is needed to reproduce the observed star formation radial profile. Given the range in the observational data, we cannot choose one of the models as a better fit, and we note the surprising similarities between the results from the PPS-L08 data set and KBMF08 data set. These simulations were run entirely independently by different investigators with unrelated initial conditions and different resolutions, and yet they produce very similar results in terms of their accretion profile in our models.

\begin{figure*}
\begin{center}
\includegraphics[scale=.70, angle=0]{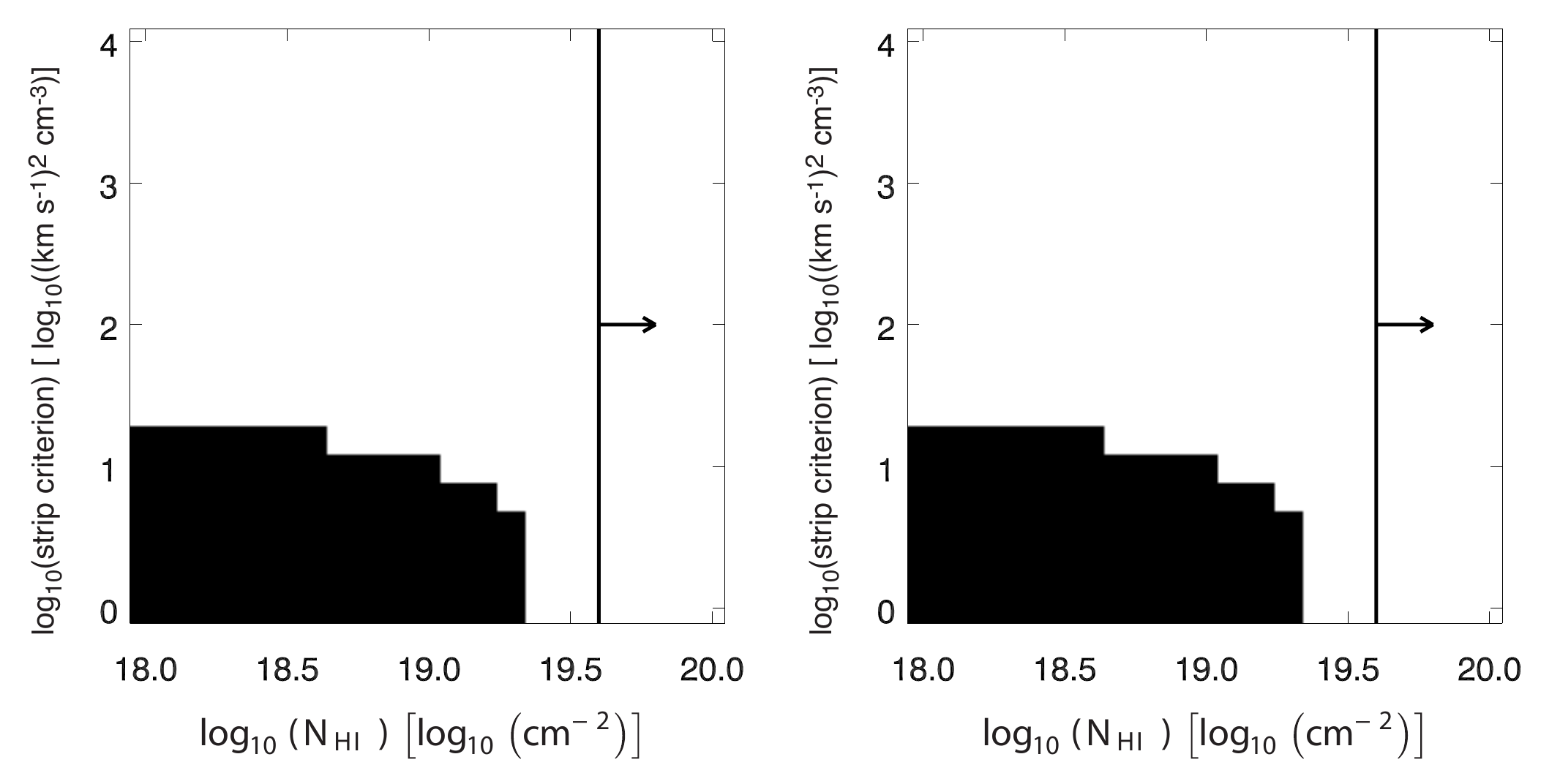}
\caption{Two plots of a large number of dwarf galaxy accretion model runs, the left plot representing runs with a low-mass halo, the right plot representing runs with a high-mass halo. Each pixel represents a run with $10^4$ dwarf galaxy objects; we vary the the column density of accreting gas ($N_H$) on the x-axis and the stripping criterion of the gas on the y-axis. Dark pixels represent model runs that meet the weak no-flow criterion ($a_{\rm out} < 0.44$). We examine a much larger range in $N_H$ than we consider observably reasonable in order to determine our sensitivity to this assumption. We also show our lower limit on typical $N_H$ with the vertical line in each plot (see \S \ref{sgstruct}). This plot demonstrates that for dwarf galaxy accretion to meet the weak no-flow criterion, typical gas column density needs to be unreasonably low.}
\label{weakgrid}
\end{center}
\end{figure*}

\begin{figure*}
\begin{center}
\includegraphics[scale=.70, angle=0]{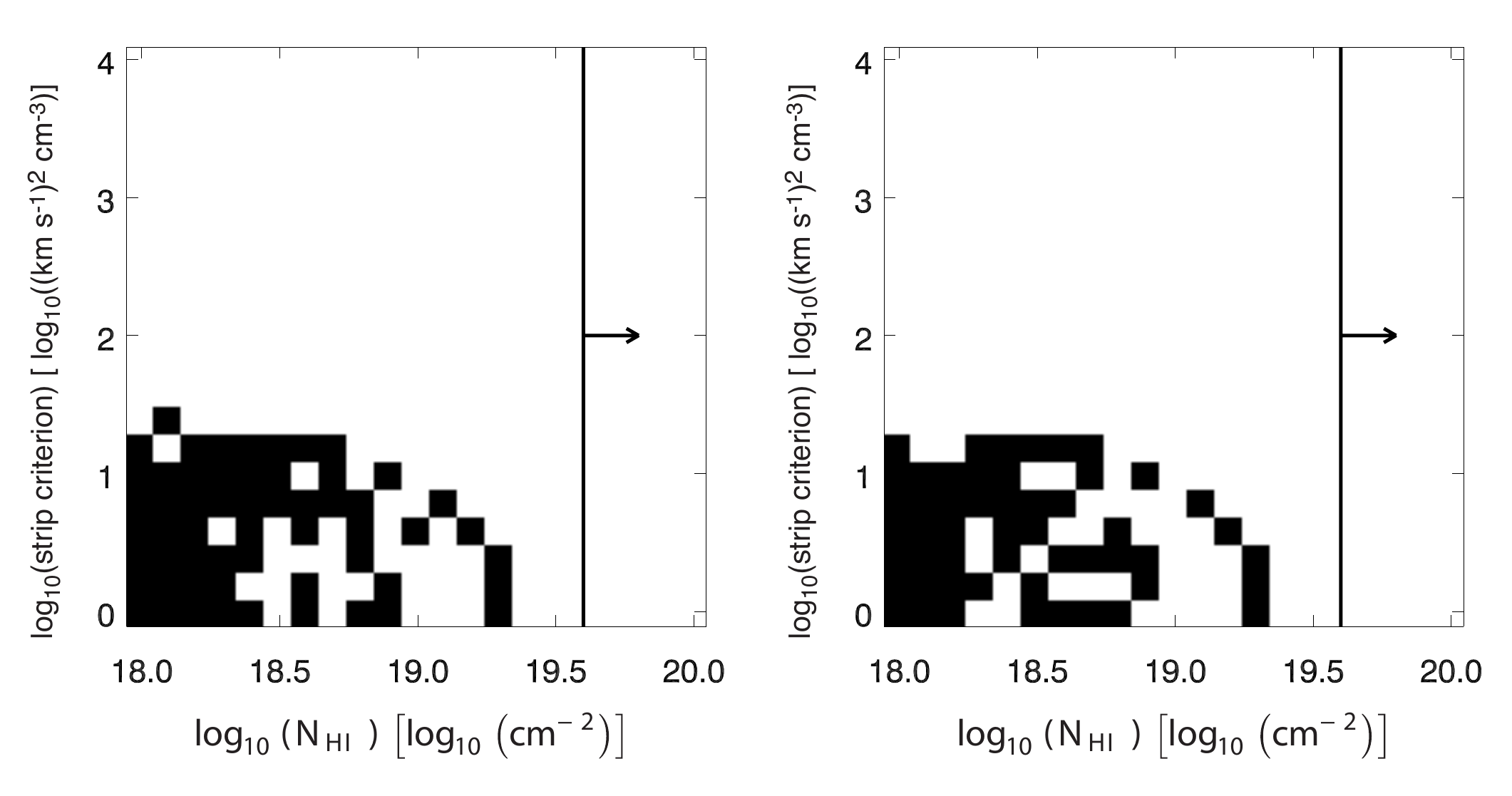}
\caption{Two plots of a large number of dwarf galaxy accretion model runs, the left plot representing runs with a low-mass halo, the right plot representing runs with a high-mass halo. Each pixel represents a model run with $10^4$ dwarf galaxy objects; we vary the the column density of accreting gas ($N_H$) on the x-axis and the stripping criterion of the gas on the y-axis. Dark pixels represent models that meet the strong no-flow criterion ($-0.35 {\rm~kpc}^{-1} \le \beta \le -0.18 {\rm~kpc}^{-1}$). We examine a much larger range in $N_H$ than we consider observably reasonable in order to determine our sensitivity to this assumptions. We also show our expected lower limit on typical $N_H$ with the vertical line in each plot (see \S \ref{sgstruct}). This plot demonstrates that for dwarf galaxy accretion to meet the strong no-flow criterion, typical gas column density needs to be unreasonably low.}
\label{stronggrid}
\end{center}
\end{figure*}

In Figures \ref{weakgrid} and \ref{stronggrid} we show the results of a larger number of dwarf galaxy model runs. Figure \ref{weakgrid} shows whether models meet the weak no-flow criterion ($a_{\rm out} < 0.44$), Figure \ref{stronggrid} shows whether models meet the strong no-flow criterion ($-0.35 {\rm~kpc}^{-1} \le \beta \le -0.18 {\rm~kpc}^{-1}$). The plots show quite clearly that, for both haloes, only runs that have unreasonably small values of $N_H$ and $\zeta < 20$ (\kmsa)$^2$ cm$^{-3}$ can meet either of the no-flow criteria.

\begin{figure*}
\begin{center}
\includegraphics[scale=.70, angle=0]{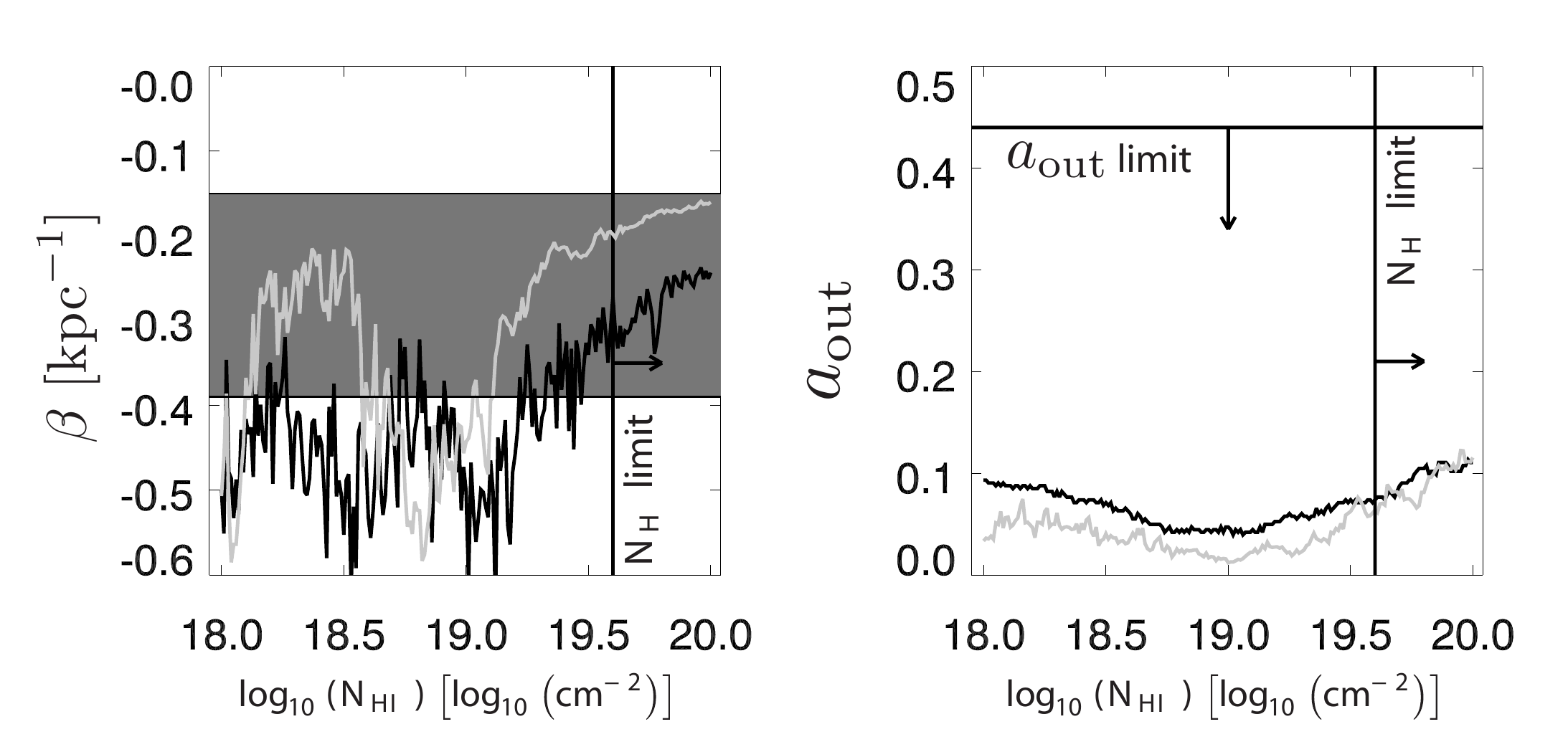}
\caption{The two plots are of $\beta$ at left and $a_{\rm out}$ at right as a function of gas column density ($N_H$) for the data taken from the cooling halo simulations (PPS-L08 in gray, KBMF08 in black). The gray box in the left plot represents the range in $\beta$ found in star-formation tracers (the strong stay put criterion), the line at the top of the right plot shows the maximum allowable value of $a_{\rm out}$ consistent with the weak no-flowno-flow criterion. The lower limit for $N_H$ determined in \S \ref{sgstruct} is also shown as a line at the right of each plot.}
\label{coolhaloplot}
\end{center}
\end{figure*}

In Figure \ref{coolhaloplot}, we show the equivalent results for the cooling halo models. Here, the stripping criterion, $\zeta$, is irrelevant (cooling clouds are not associated with stars or DM haloes), so we plot both $\beta$ (left) and $a_{\rm out}$ (right) for each model, using both the  PPS-L08 and KBFM08 simulations. We find that both the PPS-L08 and KBFM08 simulations give consistent results in $a_{\rm out}$ for all values of $N_H$ and reasonable results for $\beta$ for many values of $N_H$. For unreasonably low values of $N_H$, the cooling halo clouds tend to have too steep an accretion profile ($\beta < -0.35$). 

The sharply differing results between dwarfs and cooled halo clouds are not surprising when considering the absolute specific angular momentum, $|l|$, of the initial objects. In both the PPS-L08 and KBFM08 simulations the $\langle |l| \rangle$ of the clouds is $\sim 2.5 \times 10^3$ \kms kpc, whereas the dwarf galaxies from \citet{Benson05} had  $\langle |l| \rangle \simeq 2.5 \times 10^4$ \kms kpc. The gas in dwarf galaxies had vastly more $|l|$ to get rid of before reaching the disk, much increasing the chances of hitting the outside of the disk before reaching the center. 

\section{Conclusions}\label{concl}

We conclude from this work that the accretion of gas from dwarf galaxies cannot be the primary source of fuel for star formation in the MW without significant inward movement of gas throughout the Galactic plane. We also conclude that simulations of material cooling out of the Galactic halo naturally generate a population of clouds that is consistent with the radial profile of star formation from $R = 4$ to $R = 15$ kpc. We find that a natural distinction between these two accretion mechanisms is their typical specific angular momentum.

We hope this work will spur investigation into the fundamental parameters of ongoing Galactic accretion. One important question this work raises is whether halo drag on accreting cold gas can be modeled with the simple ram-pressure drag formulation expressed in Equation \ref{drag}. The complex interface between low- and high-density gas has bedeviled simulations (e.g., \citealt{Agertz06}), and may produce different drag effects than we have modeled. If the drag were significantly stronger our first conclusion could be incorrect, although modern grid-based simulations seem to find $C_d < 1$ (Heitsch et al, in prep). A concerted effort to determine whether ram-pressure drag is a relatively good assumption for neutral clouds in the halo and what an appropriate choice for $C_d$ is in this context could also be of use to studies of HVCs (e.g., \citealt{Peek07}). Another question raised by this work is whether all cooling halo models generate results consistent with those found from PPS-L08 and KBMF08. While no parameters were tuned in either of these simulations to match the star formation data, we cannot claim conclusively that all cooling halo models will reproduce these results, and further investigation by other groups is needed. Along these same lines, the overall accretion rate from these simulations must be examined in more detail. The accretion rate in PPS-L08 is only $\sim 0.2 M_\odot$/yr, significantly below that of the observed MW rate. It is currently unknown whether that result is a robust prediction of all cooling-halo models, but if it were confirmed it would cast doubt on the ability of a cooling halo scenario to fuel MW star formation.

The author would like to thank Tobias Kaufmann and Jesper Sommer-Larsen for supplying simulation data for the cooling halo models, Eugene Chiang for the calculation in \S \ref{radmot}, and Mary Putman, Karin Sandstrom, Steve Stahler, Louis-Benoit Desroches and Kathryn Peek for helpful comments. The author was supported in part by The National Science Foundation grants AST-0709347 and AST-0406987.

\bibliographystyle{apj}
%\bibliography{/Users/goldston/Documents/publications/mybib}

\end{document}